\documentclass[12pt]{article}

\usepackage{amsmath, mathtools}
\usepackage{geometry}
\usepackage{amssymb}
\usepackage{amsmath}
\usepackage{multirow}
\usepackage{slashed}
\usepackage{setspace}
\usepackage{fancyhdr}
\usepackage{afterpage}
\usepackage{fourier-orns}
\usepackage[mathcal]{euscript}
\usepackage{indentfirst}
\usepackage{hyperref}
\usepackage{etoolbox}
\usepackage{tikz}

\renewcommand{\=}{~=~}
\newcommand{\spartial}{\slashed{\partial}}
\newcommand{\lbar}[1]{\mkern 1.0mu\overline{\mkern-1.0mu#1\mkern-1.0mu}\mkern 1.0mu}
\newcommand{\D}{\mathcal{D}}
\newcommand{\R}{\mathbb{R}}
\newcommand{\W}{\mathcal{W}}
\renewcommand{\L}{\mathcal{L}}
\newcommand{\K}{\mathcal{K}}
\newcommand{\Z}{\mathcal{Z}}
\newcommand{\B}{\mathcal{B}}
\renewcommand{\H}{\mathcal{H}}
\newcommand{\T}{\mathcal{T}}
\renewcommand{\d}{{\rm d}}
\newcommand{\ah}{{\hat{\alpha}}}
\newcommand{\bh}{{\hat{\beta}}}
\newcommand{\gh}{{\hat{\gamma}}}
\renewcommand{\dh}{{\hat{\delta}}}
\newcommand{\ha}{{\hat{a}}}
\newcommand{\hb}{{\hat{b}}}
\newcommand{\hc}{{\hat{c}}}
\newcommand{\hd}{{\hat{d}}}
\newcommand{\he}{{\hat{e}}}
\newcommand{\ad}{{\dot{\alpha}}}
\renewcommand{\epsilon}{\varepsilon}
\newcommand{\thesistitle}{The Structure of Superforms}
\newcommand{\janksub}[1]{{s \ldots s #1 \psi \ldots \psi}}
\newcommand{\janksubp}{\janksub{}}
\newcommand{\janksubg}{\janksub{\gamma(s,\, s)}}
\newcommand{\janksubG}{\janksub{\Gamma(s,\, s)}}
\newcommand{\thoughtbreak}{\begin{center}
* \quad * \quad *\end{center}}
\newtoggle{eprint}
\toggletrue{eprint} 
\iftoggle{eprint}{
	\newcommand{\srrefcolor}{cyan}
	\newcommand{\srurlcolor}{magenta}
}{
	\newcommand{\srrefcolor}{black}
	\newcommand{\srurlcolor}{black}
}
\numberwithin{equation}{section}

\hypersetup{
	colorlinks = true,
	citecolor = \srrefcolor,
	linkcolor = \srrefcolor,
	urlcolor = \srurlcolor,
	bookmarksopen = true,
	bookmarksopenlevel = \maxdimen
}
\begin{document}


\begin{titlepage}
\begin{center}

\textsc{\LARGE University of Maryland} \\[1.5cm]
\textsc{\Large Undergraduate Thesis} \\[0.5cm]

\rule{\linewidth}{0.5mm} \\[0.4cm]
{\LARGE \textit{\thesistitle}}
\rule{\linewidth}{0.5mm} \\[1.5cm]

\begin{minipage}{0.4\textwidth}
	\begin{flushleft} \large
		\textit{Author}
		
		Stephen \textsc{Randall}
	\end{flushleft}
\end{minipage}
\begin{minipage}{0.4\textwidth}
	\begin{flushright} \large
		\textit{Advisor}
		
		Dr. William D. \textsc{Linch iii}
	\end{flushright}
\end{minipage}
\vspace{4cm}

\begin{center}
	{\large \textsc{Committee Members}}
	\vspace{0.4cm}
	
	\begin{minipage}{0.7\textwidth}
	\hrulefill	
	\vspace{-13pt}
		
	\hrulefill
		\begin{center} \large
			Dr. S. James \textsc{Gates jr} \\[4pt]
			Dr. Theodore \textsc{Jacobson} \\[4pt]
			Dr. Steven \textsc{Anlage}
		\end{center}
	\vspace{-6pt}
	\hrulefill
	\vspace{-13pt}
		
	\hrulefill
	\end{minipage}
\end{center}
 
\thispagestyle{fancy}
\fancyfoot[C]{\large \textit{December 2014}}

\end{center}
\end{titlepage}

\newpage
\iftoggle{eprint}{}{\mbox{}\thispagestyle{empty}\clearpage}

\begin{center}
	{\huge{\textit{Abstract}}}
    \vspace{24pt}
    
    {\large \textbf{\thesistitle}}
    \vspace{18pt}

    {\large Stephen Randall}
    \vspace{12pt}
    
    \textit{Center for String and Particle Theory} \\
	\textit{Department of Physics, University of Maryland}
\end{center}
\vspace{1cm}

\begin{spacing}{1.3}
	In this thesis we examine a set of foundational questions concerning closed forms in superspace. By reformulating a number of definitions through the use of a new ring of (anti-)commuting variables and the concept of an exact Bianchi form, we demonstrate a significantly streamlined method for analyzing superforms. We also study the dimensional reduction of superforms and how the relative cohomology of the superspaces involved allows for the construction of additional closed forms not in the main complex. In particular, the entire de Rham complex of closed superforms in five-dimensional superspace with eight supercharges ($N = 1$) is derived from the complex in the corresponding six-dimensional superspace. As a concluding effort, we work out the component formulation for the matter multiplets defined by five-dimensional $p$-form field-strengths for $p = 2,\, 3,\, 4$. The first and last of these come directly from the de Rham complex and coincide with multiplets that are already well-known, while the 3-form field-strength multiplet happens to require additional effort to find. This leads to the conclusion that, in general, the super-de Rham complex is not the result of supersymmetrizing the bosonic de Rham complex.
\end{spacing}

\thispagestyle{fancy}
\fancyfoot[C]{}

\newpage
\iftoggle{eprint}{}{\mbox{}\thispagestyle{empty}\clearpage}

\begin{center}
    {\huge{\textit{Acknowledgments}}}
\end{center}
\vspace{1cm}

\begin{spacing}{1.3}
	Throughout my undergraduate education I have received an incredible amount of support from an amazing ensemble of professors, friends, and family. I would not be where, nor who, I am today without them.
	
	First and foremost I thank my parents for their unconditional love and support over the past two decades. They gave me the courage to follow my passion for physics and are perhaps the only two people who care more about my future than I do. I also thank my sister Lindsay for being my oldest friend and a source of endless humor.
	
	As I began to explore my interest in theoretical physics, Drs.\ James Gates and Kory Stiffler welcomed me with open arms. They were wonderful teachers and colleagues, always taking time to carefully answer my barrages of questions. Both went above and beyond what was required of them in their efforts to further my education and I thank them wholeheartedly for the hours upon hours they invested in me.
	
	I also owe a tremendous debt of gratitude to my advisor and friend, Dr.\ William Linch. Over the past year, he has patiently taught me a staggering amount about superspace and other topics in high-energy physics. He instilled in me the maxim that everything is obvious if you think about it long enough and consistently pushed me to make sure that I understood our work to that kind of depth. His guidance, support, and lunchtime insights will be sorely missed after I leave Maryland.
	
	Additionally, I was fortunate enough to have spent four years studying with a fantastic cohort of students and friends. To Julie and Austin, thank you for those late nights our first semester in which we pushed ourselves to finish one impossible problem set after another; I would not have been able to do it alone. And to Alec, thank you for the years we spent doing homework together and for every time you challenged me to explain something that I clearly did not understand. 
	
	My final and deepest thanks go to Sarah Brown for her unending love and encouragement. Words cannot express how much you mean to me.
\end{spacing}

\thispagestyle{fancy}
\fancyfoot[C]{}

\newpage
\iftoggle{eprint}{}{\mbox{}\thispagestyle{empty}\clearpage}

\begin{spacing}{1.3}
\pagestyle{empty}
\pagenumbering{gobble}
\hypertarget{toc}{}
\noindent \hrulefill
\vspace*{-0.85cm}
\renewcommand*\contentsname{{\large Contents}}
\tableofcontents
\vspace*{-0.15cm}
\noindent \hrulefill
\end{spacing}
\newpage

\pagestyle{fancy}
\fancyfoot[C]{--- ~~\hyperlink{toc}{\thepage}~~ ---}
\pagenumbering{arabic}

\newpage
\iftoggle{eprint}{}{\mbox{}\thispagestyle{empty}\clearpage}
\setcounter{page}{1}
\section{Introduction}
\setcounter{equation}{0}
\begin{spacing}{1.3}

Originally pioneered in \cite{wes_zum, sohnius_bi, gates} and standardized as textbook material in \cite{wes_bag, 1001}, the study of closed differential forms in superspace has long enjoyed success as a source of new insights into supergravity and supersymmetric gauge theories. Partial results for different superforms in various superspaces are scattered throughout the literature\footnote{For a sampling of results from four and five dimensions, see \cite{bis_sie, hpsc, kuz_nov}.} and recently a comprehensive analysis of the de Rham complex in simple six-dimensional superspace was completed \cite{alr}. This complex of differential forms is particularly noteworthy because the closed, six-dimensional 3-form plays a key role in the famous $(2,\, 0)$ theory as a self-dual field-strength. Unfortunately, there are numerous problems associated with writing down the action for a theory containing such a field. Some of these were solved in the bosonic case \cite{per_sch} by using dimensional reduction to split the six-dimensional 3-form into two five-dimensional forms (of degree 2 and 3) that were written together in an action yielding the six-dimensional self-duality condition only as an equation of motion. However, this construction was never made supersymmetric and a proper superspace description is not obvious.

Such stumbling blocks are unsurprising given that the geometry of superspace is not universally well-understood. For example, as described in \S\ref{sec:rel_coh} the super-de Rham complex $\Omega^\bullet(\R^{m \vert n})$, surprisingly, does \textit{not} necessarily consist entirely of irreducible $p$-form supermultiplets. This generic defect impedes any attempts to describe an off-shell tensor multiplet in five or six dimensions without entering harmonic superspace. More broadly, the study of superforms itself is often not as straightforward as one might hope. In particular, there is a distinct lack of generally applicable theorems and so any new superform must be laboriously worked out from scratch. To streamline and clarify this procedure, we begin by re-examining precisely what it means for a superform to be closed and then show that the bulk of the work normally required to investigate a superform is completely unnecessary. The problem is instead reduced to the simple matter of algebraically deriving a few key Lorentz-irreducible invariants in whatever superspace is under consideration. Once these are known, the components and constraints can simply be read off from the unsolved Bianchi identities.

The outline of this thesis is as follows. We begin by reviewing the usual construction of closed superforms in \S\ref{sec:bi_usual} and demonstrate how the conventional procedure works for a five-dimensional 2-form field-strength. Continuing with this example, we show in \S\ref{sec:bi_jank} precisely how almost all of the calculations involved can be made obsolete if we move to a cleaner index-free notation and expand upon the concept and utility of a Bianchi form. The explicit demonstrations in this section require the computation of a single Lorentz-irreducible combination in $\R^{5 \vert 8}$, a process we carry out in detail in \S\ref{sec:lorentz_irred} after defining a special subset of superspaces with particularly nice spinor structure. Changing gears slightly in \S\ref{sec:dim_red}, we examine how to obtain forms in a particular superspace from those in a superspace of higher dimension but with the same number of supercharges. The dimensional reduction involved is explained in detail and an interesting source of closed superforms that are not in the main super-de Rham complex is found to arise from the relative cohomology of the two superspaces. Finally, in \S\ref{sec:5d_content} we write down the field content and component actions for the five-dimensional $p$-form field-strength supermultiplets for $p = 2,\, 3,\, 4$ and demonstrate the role of relative cohomology in finding the (on-shell) tensor multiplet in five dimensions. Almost all examples we present herein are set in the context of five-dimensional superspace with eight supercharges and our conventions for this space are laid out in \S\ref{sec:susy_math}.

\section{Conventional Superforms}
\label{sec:bi_usual}

A $p$-form $\omega \in \Omega^p(\R^{m \vert n})$ can be expressed in terms of the superspace basis co-tangent vectors $\d z^M$ as
\begin{equation}
	\omega \= \frac{1}{p!} \d z^{M_1} \wedge \ldots \wedge \d z^{M_p} \omega_{M_p \ldots M_1} \,,
\end{equation}
where $z^M = (x^a,\, \theta^\alpha)$ and the components $\omega_{M_p \ldots M_1}$ are organized by number of spinor indices. Since the differential $\d = \d z^M \partial_M$ does not commute with the supersymmetry generators, we instead change bases in the standard fashion \textit{via} the framing
\begin{equation}
	e^A \= \d z^M e_M{}^A \,,
\end{equation}
chosen so that the differential becomes
\begin{equation}
	\d \= \d z^M \partial_M \= e^A \D_A
\end{equation}
and the $\D_A = (\D_\alpha,\, \partial_a)$ are super-covariant derivatives. The price to be paid for this change is the emergence of torsion from the requirement that the differential be nilpotent of order 2 while $[\D_A,\, \D_B] = f_{AB}{}^C \D_C$ with $f_{\alpha \beta}^a$ in particular always non-vanishing. This torsion is
\begin{equation}
\label{eq:conv_torsion}
	T^A \= \d e^A \= \d z^M \d z^N T_{MN}{}^A \,,
\end{equation}
where
\begin{equation}
	T_{MN}{}^C \= e_N{}^A e_M{}^B f_{AB}{}^C \,.
\end{equation}
The $p$-form $\omega$ is then re-written as
\begin{equation}
	\omega \= \frac{1}{p!} e^{A_1} \wedge \ldots \wedge e^{A_p} \omega_{A_p \ldots A_1}
\end{equation}
and its differential can be put in the form
\begin{equation}
	\d \omega \= \frac{1}{(p + 1)!} e^{A_1} \wedge \ldots \wedge e^{A_{p + 1}} B(\omega)_{A_{p + 1} \ldots A_1} \,.
\end{equation}
The components of $B(\omega)$ are then
\begin{equation}
\label{eq:conv_bi}
	B(\omega)_{A_1 \ldots A_{p + 1}} \= \frac{(p + 1)!}{p!} \D_{[A_1} \omega_{A_2 \ldots A_{p + 1}]} + \frac{(p + 1)!}{2!(p - 1)!} T_{[A_1 A_2 \vert}{}^C \omega_{C \vert A_3 \ldots A_{p + 1}]} \,,
	\vspace*{-0.05cm}
\end{equation}
\makebox[0.985\linewidth][s]{where $[\cdot]$ denotes the graded anti-symmetric part of the enclosed index structure.}\footnote{That is, (composite) spinor indices are symmetrized while vector indices are anti-symmetrized.}
\clearpage

\noindent We will refer to the exact $(p + 1)$-form $B$ as a \textit{Bianchi form}. Noting that $\omega$ is closed if and only if all the components of $B(\omega)$ vanish, it will be important later that $B(B(\omega)) \equiv 0$ actually gives a large amount of information about the structure of $\omega$.

The general utility of superforms comes from their natural accommodation of gauge structure. Analogously to the bosonic construction, if $A$ is an abelian gauge $(p - 1)$-form then its field-strength $F$ is simply defined as the $p$-form
\begin{equation}
	F \= \d A \,.
\end{equation}
This field-strength is invariant under the gauge transformation $\delta A = \d \lambda$ for any $\lambda \in \Omega^{p - 2}(\R^{m \vert n})$ and is itself identically closed. As a superform, $F$ is expected to describe a constrained, gauge-invariant superfield whose $\theta$-expansion holds the field content of a $(p - 1)$-form gauge supermultiplet. To find this superfield and the constraints it satisfies, we are required to solve the closure conditions\footnote{These conditions are also commonly referred to as Bianchi identities.}
\begin{equation}
\label{eq:conv_closure}
	0 \= \frac{1}{p!} \D_{[A_1} F_{A_2 \ldots A_{p + 1}]} + \frac{1}{2!(p - 1)!} T_{[A_1 A_2 \vert}{}^C F_{C \vert A_3 \ldots A_{p + 1}]} \,.
\end{equation}
For an explicit example, we specialize to $m \vert n = 5 \vert 8$ and $p = 2$ in the next section.

\subsection{A Five-Dimensional 2-form}
\label{sec:5d_2form_vanilla}

Consider the five-dimensional superspace with eight supercharges, our conventions for which are reviewed in \S\ref{sec:susy_math}. The 2-form field-strength $F \in \Omega^2(\R^{5 \vert 8})$ describes a Yang-Mills supermultiplet when it is exact as $F = \d A$ and expressed in terms of a scalar superfield. We omit the analysis of $A$ for the sake of brevity but it easy to find that the closure of a generic 1-form $A$ requires the imposition of the scalar constraint
\begin{equation}
	\D^{\ah i} A_{\ah i} \= 0
\end{equation}
\makebox[\linewidth][s]{on the spin component $A_{\ah i}$ of $A$ at dimension $1$ in the closure conditions. In order to}
\clearpage

\noindent have $F = \d A$ we then set the lowest component of $F$ to be the obstruction to this condition. That is, we define
\begin{equation}
	F_{\ah i \bh j} \= B(A)_{\ah i \bh j} ~=:~ 2i \epsilon_{ij} \epsilon_{\ah \bh} \W \,,
\end{equation}
for some dimension-1 field-strength $\W$. Now that we have the lowest component of $F$, the remaining components and any constraints on $\W$ follow uniquely from \eqref{eq:conv_closure}.

To begin, consider the dimension-$\tfrac{3}{2}$ condition $0 = B(F)_{\ah i \bh j \gh k}$ which becomes
\begin{equation}
	0 \= \D_{\ah i} F_{\bh j \gh k} + 2i \epsilon_{ij} (\Gamma^\ha)_{\ah \bh} F_{\gh k \ha} + (\underline{\alpha \beta \gamma}) \,.
\end{equation}
Here $\underline{\alpha} \equiv \ah i$ and the notation $(\underline{\,\cdot\,})$ denotes the remaining cyclic permutations of the enclosed composite indices. Plugging in $F_{\ah i \bh j}$, we find that $F_{\ah i \ha}$ is fixed to be
\begin{equation}
	F_{\ah i \ha} \= - (\Gamma_\ha)_\ah{}^\bh \D_{\bh i} \W \,.
\end{equation}
The dimension-2 condition $0 = B(F)_{\ah i \bh j \ha}$ becomes
\begin{equation}
	0 \= 2 \D_{\ah i} F_{\bh j \ha} + \partial_\ha F_{\ah i \bh j} - 2i \epsilon_{ij} (\Gamma^\hb)_{\ah \bh} F_{\hb \ha} + (\underline{\alpha \beta}) \,,
\end{equation}
and upon plugging in the known components, we have
\begin{equation}
\label{eq:2form_van_dim2}
	0 \= [ (\Gamma_\ha)_\bh{}^\gh \D_{\ah i} \D_{\gh j} + (\Gamma_\ha)_\ah{}^\gh \D_{\bh j} \D_{\gh i} ] \W - 2 i \epsilon_{ij} \epsilon_{\ah \bh} \partial_\ha \W + 2i \epsilon_{ij} (\Gamma^\hb)_{\ah \bh} F_{\hb \ha} \,.
\end{equation}
To solve this, we expand the $\D \D$ parts using \eqref{eq:dd_exp} and plug back into \eqref{eq:2form_van_dim2}, yielding
\begin{align}
\label{eq:2form_van_dim2_exp}
	0 & \= [ - i \epsilon_{ij} (\Gamma_\ha \Gamma^\hb)_{\bh \ah} \partial_\hb - \frac{1}{2} \epsilon_{ij} (\Gamma_\ha \Sigma^{\hb \hc})_{\bh \ah} \D^2_{\hb \hc} + \frac{1}{2} (\Gamma_\ha \Gamma^\hb)_{\bh \ah} \D^2_{\hb i j} - \frac{1}{2} (\Gamma_\ha)_{\bh \ah} \D^2_{ij} \notag\\[4pt]
		& \quad +\, (\underline{\alpha \beta}) ] \W - 2 i \epsilon_{ij} \epsilon_{\ah \bh} \partial_\ha \W + 2i \epsilon_{ij} (\Gamma^\hb)_{\ah \bh} F_{\hb \ha} \,.
\end{align}
The $(\underline{\alpha \beta})$ symmetry kills the final term in the $\D \D$ expansion and allows the $\partial \W$ terms to cancel. Additionally, it restricts the irreducibles in the remaining two terms of the $\D \D$ expansion, leaving behind the relation
\begin{equation}
	0 \= [- \epsilon_{ij} (\Gamma^\hb)_{\bh \ah} \D^2_{\ha \hb} - 2 (\Sigma_\ha{}^\hb)_{\ah \bh} \D^2_{\hb ij}] \W + 2i \epsilon_{ij} (\Gamma^\hb)_{\ah \bh} F_{\hb \ha} \,.
\end{equation}
Because of the (anti-)symmetry in the $ij$ indices, this is actually two separate conditions; one defines $F_{\ha \hb}$ and the other puts a restriction on $\W$. The former yields
\begin{equation}
	F_{\ha \hb} \= - \frac{i}{2} \D^2_{\ha \hb} \W \,,
\end{equation}
while the latter requires
\begin{equation}
\label{eq:2form_dim2_acons}
	\D^2_{\ha ij} \W \= 0 \,.
\end{equation}
From \eqref{eq:dd_exp}, it is clear that \eqref{eq:2form_dim2_acons} is equivalent to
\begin{equation}
\label{eq:2form_avm_cons}
	\D_\ah^{(i} \D_\bh^{j)} \W \= \frac{1}{4} \epsilon_{\ah \bh} \D^{\gh (i} \D_\gh^{j)} \W \,.
\end{equation}
Continuing with the dimension-$\tfrac{5}{2}$ condition, we plug in the components of $F$ to find
\begin{equation}
\label{eq:2form_dim52_bi}
	\D_{\ah i} \D^k_{(\bh} \D_{\gh) k} \W \= 4i \spartial_{\dh (\bh} \epsilon_{\gh) \ah} \D^\dh_i \W - 4i \spartial_{\ah (\bh} \D_{\gh) i} \W \,.
\end{equation}
Through a bit of $\Gamma$-matrix algebra this can be shown to come directly from \eqref{eq:2form_dim2_acons} by expanding and simplifying
\begin{equation}
\label{eq:2form_dim52_check}
	(\Gamma_\ha)_{\ah \bh} (\Gamma_\hb)_{\gh \dh} (\Sigma^{\ha \hb})_{\hat{\rho} \hat{\tau}} \D^{\bh i} \D^\gh_{(i} \D^\dh_{j)} \W \= 0 \,.
\end{equation}
The dimension-3 closure condition is the one familiar from the study of differential forms in bosonic space,
\begin{equation}
\label{eq:2form_dim3_bi}
	\partial_{[\ha} F_{\hb \hc]} \= 0 \,,
\end{equation}
and like the dimension-$\tfrac{5}{2}$ condition \eqref{eq:2form_dim52_bi} it holds identically since
\begin{equation}
	\epsilon_{\ha \hb}{}^{\hc \hd \he} \partial_\hc F_{\hd \he} \= - \frac{i}{2} \epsilon_{\ha \hb}{}^{\hc \hd \he} \partial_\hc \D^2_{\hd \he} \W \= \frac{1}{12} [\D^2_{\ha ij},\, \D^{2 ij}_\hb] \W \= 0 \,.
\end{equation}
Thus, the only constraint on $\W$ is \eqref{eq:2form_dim2_acons}. And as we will see in \S\ref{sec:vect_mult}, this is exactly the superfield constraint that defines a five-dimensional vector multiplet.

\vspace{2pt}
\thoughtbreak
\vspace{-4pt}

In general, this procedure requires a fair amount of foresight. It is not obvious why \eqref{eq:2form_dim52_check} was the combination that needed to be expanded and simplified to verify that the dimension-$\tfrac{5}{2}$ condition \eqref{eq:2form_dim52_bi} was already satisfied. And for the dimension-3 condition \eqref{eq:2form_dim3_bi}, it was necessary to know precisely the right $\D$-identity. Additionally, the concept of generating a closed $p$-form by obstructing the closure of a $(p - 1)$-form is not always used, with lowest components sometimes simply being guessed and then arduously examined for consistency. Although this situation has not stopped anyone from using superforms to make tremendous progress in the investigation of supersymmetric gauge theories, it should be apparent that this process is more of an art than a science.

One instance of this is the common lore that the top two Bianchi identities---here the dimension-$\tfrac{5}{2}$ and 3 conditions---never impose any new constraints on the form and therefore do not really need to be examined. We have just shown this to be true for $F$ but unfortunately there is no universal way to avoid these last checks. On top of this, higher degree forms often require additional non-trivial calculations in the process of finding constraints. For example, the five-dimensional 3-form (which we will work through in \S\ref{sec:prior_cons}) has two constraints and the second is very difficult to isolate from relations already implied by the first. In the next section we address these concerns conclusively by introducing more advanced machinery that aids in the formulation of general theorems about superspace cohomology.

\section{Index-free Notation and Bianchi Forms}
\label{sec:bi_jank}

The approach outlined in the previous section is neither simple nor illuminating. There is no \textit{a priori} method for knowing precisely what needs to be computed to get to a desired result and for more complicated forms the procedure becomes heavily dependent on intuition and experience. Additionally, if the top closure conditions truly never hold any new information about a form then we should not waste time computing anything at these levels. A complete understanding of the generic structure of a superform would clarify these aspects of the standard analysis.

To that end, we now introduce a super-commutative ring over $\R$ with elements $s^M = (s^\alpha,\, \psi^a)$. The $s$-variables are commuting spinors while the $\psi$-variables are anti-commuting vectors. That is, they have the commutation relations
\begin{equation}
	s^\alpha s^\beta \= s^\beta s^\alpha \,, \qquad s^\alpha \psi^a \= \psi^a s^\alpha \,, \qquad \psi^a \psi^b \= - \psi^b \psi^a \,,
\end{equation}
which are chosen so that if we contract these variables on \eqref{eq:conv_closure} they will automatically encode the super-wedge product structure. Importantly, these variables do not carry torsion in contrast with \eqref{eq:conv_torsion}. Performing the contractions then gives a set of closure conditions in flat superspace in terms of Bianchi form components,
\begin{align}
\label{eq:jank_bi}
	0 \= B(\omega)_\janksubp & \= s \D_s \omega_\janksubp + (-1)^s (p + 1 - s) \partial_\psi \omega_\janksubp \notag\\
			& \quad - i (-1)^s s(s - 1) \omega_\janksubg \,,
\end{align}
where $s$ is the number of (spinor) $s$-contractions, $p$ is the degree of $\omega$, and we use the shorthand $X^M A_M =: A_X$. Notice that relative to \eqref{eq:conv_closure}, the Bianchi components $B(\omega)_\janksubp$ have been re-scaled by a factor of $s!$ for convenience. We have also explicitly plugged in the basic supersymmetry torsion $T_{\alpha \beta}^a \sim (\gamma^a)_{\alpha \beta}$ and defined $\gamma^a(s,\, s) := s^\alpha (\gamma^a)_{\alpha \beta} s^\beta$.\footnote{This notation is more suggestive than formally correct. In six dimensions, for example, the explicit form also has contracted $\mathsf{SU}(2)$ indices: $\gamma^a(s,\, s) = \epsilon_{ij} s^{\alpha i} (\gamma^a)_{\alpha \beta} s^{\beta j}$. In four dimensions, the spinor splits as $s \rightarrow s \oplus \lbar{s}$ and so $\gamma^a(s,\, \lbar{s}) = s^\alpha (\sigma^a)_{\alpha \ad} \lbar{s}^\ad$. The point is simply that this is the form of the flat space torsion and can be made precise upon specializing to a specific superspace.} This leads to the compact formulation of the flat space supersymmetry algebra,
\begin{equation}
	\D_s^2 \= i \partial_{\gamma(s,\, s)}.
	\vspace*{0.5cm}
\end{equation}
In curved superspace, we introduce the curved supercovariant derivatives $\mathfrak{D}_A$ and the torsions $T_{\psi \psi}^\alpha$ and $T_{s \psi}^\alpha$.\footnote{In six dimensions, these derivatives and torsions are explicitly defined in \cite{tm_lin}.} The closure conditions in curved space are then
\begin{align}
\label{eq:jank_bi_curved}
	0 & \= s \mathfrak{D}_s \omega_\janksubp + (-1)^s (p + 1 - s) \mathfrak{D}_\psi \omega_\janksubp \notag\\
		& \quad - i (-1)^s s (s - 1) \omega_\janksubg + (-1)^s s (p + 1 - s) T_{s \psi}^\alpha \omega_{\alpha \janksubp} \notag\\
		& \quad - \tfrac{1}{2} (p + 1 - s)(p - s) T_{\psi \psi}^\alpha \omega_{\alpha \janksubp} \,.
\end{align}
Unless otherwise noted we will restrict ourselves to flat space, although one of the benefits of this new formulation is how nicely it extends to curved space.

With the index structure effectively abstracted away in the closure conditions \eqref{eq:jank_bi}, we can now begin to look directly at the inner workings of superforms. As is the case in the conventional approach, the starting point for studying a closed superform is to see how many components of the form we are allowed to set to zero. This procedure primarily ensures that we obtain superfield representations that are irreducible and therefore physically interesting. From \eqref{eq:jank_bi}, note that if the lowest (with respect to mass-dimension) component $\omega_{s \ldots s}$ vanishes\footnote{While this is by far the most common scenario in super-de Rham complexes, we have already seen an example where $\omega_{s \ldots s}$ is non-zero. In \S\ref{sec:5d_2form_vanilla} we had $F_{ss} \sim s^2 \W$ as an allowed lowest component. However, this is a very special case due to the low degree of the form and so most of the following discussion will in fact be under the general assumption that $\omega_{s \ldots s} = 0$.} then the lowest non-vanishing (lnv) component of $\omega$ must satisfy the condition
\begin{equation}
\label{eq:lnv_comp}
	\omega^\text{lnv}_\janksubg \= 0 \,.
\end{equation}
It is therefore important to figure out which objects are killed under $\psi \rightarrow \gamma(s,\, s)$ contraction, as these are the main objects used to define the lowest component of a closed superform. Before doing so, let us take a moment to address the larger issue with the conventional approach: the difficulty of finding constraints.

As discussed at the end of \S\ref{sec:5d_2form_vanilla}, it is generally not straightforward to check the closure conditions (particularly the higher-dimensional ones) for constraints on the field-strength. However, we can now show that finding constraints is effectively equivalent to determining the solutions to \eqref{eq:lnv_comp}. Recall the definition of the Bianchi form as in \eqref{eq:jank_bi} and note that because the superspace differential squares to zero, we have that $B(B(\omega)) \equiv 0$. That is, $B(\omega)$ is an identically closed form already. Interestingly, the information encoded in this statement will pinpoint precisely where the constraints on $\omega$ are hiding in the closure conditions.

For a $p$-form $\omega$, we first need to solve the equation $B_{s \ldots s} = 0$ if it is not completely trivial. The magic is then that this is the \textit{only} Bianchi identity that we will need to solve explicitly and it is always the easiest one. Now by an inductive argument, assume that we have solved the Bianchi identities up through dimension-$(d - \tfrac{1}{2})$ by finding constraints on $\omega$. Because all prior components of the Bianchi form vanish, the dimension-$d$ component now identically satisfies
\begin{equation}
\label{eq:Bcons_jank}
	B_\janksubg \= 0 \,.
\end{equation}
Note the difference between \eqref{eq:Bcons_jank} and \eqref{eq:lnv_comp}: while \eqref{eq:lnv_comp} is required to hold for the lowest component of $\omega$, \eqref{eq:Bcons_jank} holds automatically for \textit{every} Bianchi component when we have solved all previous Bianchi identities due to the fact that $B$, as defined, is identically closed. That is, the fact that the lower Bianchi components have been constrained to vanish implies directly that the only remaining obstruction to the closure of $\omega$ at this level must be in the kernel of $\psi \rightarrow \gamma(s,\, s)$ contraction.

Since we have reduced the determination of both the superform components and constraints to the same problem, let us now begin to solve it. We first want to introduce the notion of an \textit{$L$-combination} as a non-trivial Lorentz-irreducible combination of $\gamma$-matrices and $s$-variables that satisfies the condition
\begin{equation}
\label{eq:Lc_jank}
	L_\janksubg ~\equiv~ 0 \,.
\end{equation}
These combinations are fixed by the spinor structure of the superspace under consideration. Requiring no other information, they then tell us exactly what the components and constraints look like for any superform in this space. Unless there is some low-degree coincidence (such as the $p = 2$ superform in five dimensions that we examined in \S\ref{sec:5d_2form_vanilla}), the lowest component of every superform must be an $L$-combination. Once the lowest component of a form is known, the higher components follow easily from the Bianchi identities in the conventional fashion.

To see how the $L$-combinations fix the constraints as well, note that from the structure of the closure conditions \eqref{eq:jank_bi} the only objects allowed inside a Bianchi component satisfying \eqref{eq:Bcons_jank} are
\begin{equation}
\label{eq:bianchi_rem_irred}
	B_\janksubp \= \omega_\janksubg + \sum_\ell \left( L^\ell \cdot C^\ell \right)_\janksubp \,,
\end{equation}
where each $L^\ell$ is an $L$-combination and the ``$\cdot$" refers to the contraction of any remaining indices on $L$ (as in \eqref{eq:2formB_dim2_irred_exp}, for example). The first term in \eqref{eq:bianchi_rem_irred} defines the next higher component in $\omega$ by absorbing the exact part (more specifically, the $\psi \rightarrow \gamma(s,\, s)$ contraction of the exact part) of the full Bianchi component. Then because the remaining portion has nothing left to cancel against, if we are demanding that $B_\janksubp = 0$ then we are required to set each $C^\ell = 0$. These are precisely the constraints on the field-strength at this level. We will see two examples of how this process works in \S\ref{sec:5d_2form_jank} and \S\ref{sec:prior_cons} but the gist of it is that we first isolate the $L$-combinations in the relevant superspace, write down a schematic equation for the Bianchi component \textit{\`a la} \eqref{eq:bianchi_rem_irred}, and then plug in the actual superform components to determine each $C^\ell$.

Before moving on to the examples however, it is worth noting how simply this procedure generalizes to curved space. Recall the curved closure conditions \eqref{eq:jank_bi_curved} and the fact that the new torsions contract over a spinor index. This means that the components sitting behind them are of a lower dimension than the component sitting behind the flat space torsion and therefore \eqref{eq:Bcons_jank} is unchanged. So although the components and constraints will be more complicated, no part of the approach itself actually needs to change to accommodate curved superspaces.

\subsection{Re-examining the 2-form}
\label{sec:5d_2form_jank}

Using this technology, we can now re-analyze the five-dimensional 2-form $F$. Again we set $F_{ss} = 2i s^2 \W$ and so the full set of components is uniquely fixed to
\begin{align}
\label{eq:2form_comp_jank}
	F_{ss} & \= 2i s^2 \W \,, \notag\\[4pt]
	F_{s \psi} & \= - s^i \Gamma_\psi \D_i \W \,, \notag\\[4pt]
	F_{\psi \psi} & \= - \frac{i}{2} \D^2_{\psi \psi} \W \,,
\end{align}
equivalent to the set in \S\ref{sec:5d_2form_vanilla}. It is important to note that we are not simply guessing these components and then checking their consistency. The derivation of the higher components from the lowest non-vanshing one is never difficult and in flat space simply consists of pulling off the $\gamma(s,\, s)$ contraction (cf. \S\ref{sec:5d_2form_vanilla}). Since we wish to focus primarily on finding constraints, we purposefully omit this part of the analysis.

Before moving on, we need to understand the $L$-combinations in this superspace. We discuss how these are obtained in \S\ref{sec:lorentz_irred} but the defining relation in $\R^{5 \vert 8}$ is
\begin{equation}
\label{eq:li_def_5d}
	\Gamma^\ha(s,\, s) \Sigma_{\ha \hb}(s^i,\, s^j) \= 0 \,,
\end{equation}
and so the only irreducible combination of $\Gamma$-matrices in five dimensions killed under $\psi \rightarrow \Gamma(s,\, s)$ contraction is $L_{ss \psi \psi}^{ij} = \Sigma_{\psi \psi}(s^i,\, s^j)$. It is worth reiterating that this object alone tells us a significant amount about the structure of five-dimensional superspace and is essentially all we need to work out the entire complex of superforms in five dimensions. Let us now see exactly how this $L$-combination is utilized and why we are able to make such a claim about its usefulness.

The Bianchi form $B = B(F)$ first satisfies the dimension-2 Bianchi identity
\begin{equation}
\label{eq:2formB_dim2}
	0 \= 4 \D_s B_{sss} + 12i B_{ss \Gamma(s,\, s)} \,.
\end{equation}
Since $B_{sss}$ vanishes algebraically from the form of the components and requires no differential constraints on $\W$, \eqref{eq:2formB_dim2} becomes
\begin{equation}
	B_{ss \Gamma(s,\, s)} \= 0 \,.
\end{equation}
The only possible terms in $B_{ss \psi}$ are then
\begin{equation}
\label{eq:2formB_dim2_irred_exp}
	B_{ss \psi} ~\sim~ F_{\Gamma(s,\, s) \psi} + s^i \Sigma_\psi{}^\ha s^j C_{\ha ij} \,.
\end{equation}
Comparing back to \eqref{eq:bianchi_rem_irred}, the first term in \eqref{eq:2formB_dim2_irred_exp} defines the next higher component, while what is left can only be interpreted as a constraint on $\W$. If we write out the explicit form of $B_{ss \psi}$ in terms of the components of $F$, we have
\begin{equation}
	B_{ss \psi} \= - 2 s^i \D_s \Gamma_\psi \D_i \W + 2i s^2 \partial_\psi \W + 2i F_{\Gamma(s,\, s) \psi} \,.
\end{equation}
The third term is uninteresting after we have used it to determine $F_{\psi \psi}$, so we focus on extracting from the first two terms the constraint piece in \eqref{eq:2formB_dim2_irred_exp}. Only the $\D \Gamma \D$ term is still reducible and using \eqref{eq:dd_exp} with \eqref{eq:gab_gamma_gamma} we find that the vector $\D^2$ irreducible plays the role of $C$ here. That is, the dimension-2 constraint on $\W$ can---with no more than very basic knowledge of the $\Gamma$-matrix algebra in five dimensions---be directly read off as
\begin{equation}
\label{eq:2form_vcons}
	0 \= C_{\ha ij} \= \D^2_{\ha ij} \W \,.
\end{equation}
The remaining portions of $B_{ss \psi}$ will cancel against each other, as required by the component structure of $F$ that allowed $B_{sss}$ to vanish identically. Notice that we did not have to worry about any of the other terms that were involved in the expansion \eqref{eq:2form_van_dim2_exp}; we were able to pull out precisely the term we were looking for and nothing else. As we study more complicated forms higher in the complex, this ability will prove to be quite valuable.

The final two Bianchi identities are now very simple. Neither $B_{s \psi \psi}$ nor $B_{\psi \psi \psi}$ defines a new component and there are no Lorentz-irreducible combinations in the kernel of $\psi \rightarrow \Gamma(s,\, s)$ contraction with the correct index structure to appear in these Bianchi components. That means that there is no room for any new constraints in the top two Bianchi identities and we have therefore finished the analysis for the 2-form in five dimensions. Contrast this with the examination of the top two Bianchi identities in \S\ref{sec:5d_2form_vanilla}: the dimension-$\tfrac{5}{2}$ condition required that we guess precisely the right object to compute while verification of the dimension-3 condition needed just the right $\D$-identity. So with a little bit of front-loading in terms of figuring out the $L$-combinations of $\R^{5 \vert 8}$, the actual analysis of every superform in this superspace has become completely algorithmic. One nice result that we can already prove is that the top two Bianchi identities imply no new constraints on \textit{any} form in this superspace: because the only $L$-combination here is defined by \eqref{eq:li_def_5d} and requires two $s$-variables, it is impossible to write any constraint term for $B_{s \psi \ldots \psi}$ or $B_{\psi \ldots \psi}$.

In the next section, we continue up the complex to the 3-form $H$. This form is considerably more complicated than the 2-form due to the fact that there are two constraints involved. As mentioned at the end of \S\ref{sec:5d_2form_vanilla}, the usual issue with multiple constraints is ensuring that the second is not already implied by the first. But as we will see, this approach allows us to straightforwardly isolate all \textit{independent} constraints on the field-strength with far less effort than is usually required.

\subsection{Using Prior Constraints}
\label{sec:prior_cons}

Consider the five-dimensional 3-form $H$ defined by obstructing the closure of $F$ with the dimension-1 field-strength $H_{\ha ij} = C_{\ha ij}$ in \eqref{eq:2form_vcons}. This gives $H_{sss} = 0$ and so the components of this form are worked out to be
\vspace*{-0.1cm}
\begin{align}
	H_{sss} & \= 0 \,, \notag\\[4pt]
	H_{ss \psi} & \= - s^i \Sigma_{\psi \ha} s^j H_{ij}^\ha \,, \notag\\[4pt]
	H_{s \psi \psi} & \= \frac{i}{12} \epsilon_{\psi \psi}{}^{\ha \hb \hc} s^i \Sigma_{\ha \hb} \D^j H_{\hc ij} \,, \notag\\[8pt]
	H_{\psi \psi \psi} & \= \frac{1}{48} \epsilon_{\psi \psi \psi}{}^{\ha \hb} \D_{\ha i j}^2 H_\hb^{ij} \,.
\end{align}
The lowest Bianchi component is identically zero due to the form of the lowest components of $H$, so we begin by noting that the form of the next higher component is restricted to be
\begin{equation}
	B_{sss \psi} ~\sim~ H_{s \Gamma(s,\, s) \psi} + s^i \Sigma_\psi{}^\ha s^j s^{\ah k} C_{\ah \ha ijk} \,.
\end{equation}
The actual form of this Bianchi component, after plugging in $H_{sss}$ and $H_{ss \psi}$ is
\begin{equation}
	B_{sss \psi} \= - 3 \D_s s^i \Sigma_{\psi \ha} s^j H_{ij}^\ha - 6i H_{s \Gamma(s,\, s) \psi} \,,
\end{equation}
and so the constraint at this level becomes
\begin{equation}
\label{eq:lower_cons}
	0 \= (\Sigma_{\ha \hb})_{(\ah \bh} \D_{\gh) (i} H_{jk)}^\hb \,.
\end{equation}
Using the $\Gamma$-traceless projection operator $\Pi_{\ha \ah}^{~ \hb \bh} := \delta_\ha^\hb \delta_\ah^\bh + \frac{1}{5} (\Gamma_\ha \Gamma^\hb)_\ah{}^\bh$ this is equivalent to the condition
\begin{equation}
\label{eq:3form_dim52_cons}
	0 \= C_{\ah \ha ijk} \= \Pi_{\ha \ah}^{~ \hb \bh} \D_{\bh (i} H_{\hb jk)} \,,
\end{equation}
which can be seen by expanding $\D H = (\Pi + \Pi^\perp) \D H$ and then acting with the appropriate $\Sigma$ (while totally symmetrizing the necessary spinor indices). The $\Pi^\perp$ part vanishes and we are left with \eqref{eq:3form_dim52_cons}.

Now with $B_{sss \psi} = 0$ solved, the next Bianchi component has the form
\begin{equation}
\label{eq:3form_ssvv_struct}
	B_{ss \psi \psi} ~\sim~ H_{\Gamma(s,\, s) \psi \psi} + s^i \Sigma_{\psi \psi} s^j C_{ij} \,.
\end{equation}
Plugging in the components of $H$, the actual Bianchi component looks like
\begin{equation}
\label{eq:3form_ssvv_actual}
	B_{ss \psi \psi} \= \frac{i}{6} \epsilon_{\psi \psi}{}^{\ha \hb \hc} \D_s s^i \Sigma_{\ha \hb} \D^j H_{\hc ij} - 2 \partial_\psi s^i \Sigma_{\psi \ha} s^j H^\ha_{ij} + 2i H_{\Gamma(s,\, s) \psi \psi} \,.
\end{equation}
The third term does not contribute to the constraint at this level and so we look only at the first two. What \eqref{eq:3form_ssvv_struct} tells us is that the constraint sits behind the $L^{ij}_{ss \psi \psi} = s^i \Sigma_{\psi \psi} s^j$ part and so we must pull that out of the first two terms in \eqref{eq:3form_ssvv_actual}. In five dimensions we note that for $T^{\ha \hb}_{\ah \bh} = T^{[\ha \hb]}_{(\ah \bh)}$, we can---through use of the $\Gamma$-matrix identities \eqref{eq:gab_trace}, \eqref{eq:gab_gamma_gamma}, and \eqref{eq:gab_gamma_sigma}---write
\begin{equation}
	T^{\ha \hb}_{\ah \bh} \= \frac{1}{20} (\Sigma^{\ha \hb})_{\ah \bh} (\Sigma_{\hc \hd})^{\gh \dh} T^{\hc \hd}_{\gh \dh} + ~\cdots~ \,.
\end{equation}
This lets us express the $\partial_\psi$ term as
\begin{equation}
	- 2 \partial_\psi s^i \Sigma_{\psi \ha} s^j H^\ha_{ij} \= \frac{2}{5} s^i \Sigma_{\psi \psi} s^j \partial_\ha H^\ha_{ij} + ~\cdots~ \,,
\end{equation}
where the terms we are ignoring do not play any part in the constraint at this level because they do not have the correct $L \cdot C$ structure. A similar but slightly lengthier calculation shows that we can also express the first term in \eqref{eq:3form_ssvv_actual} as
\begin{equation}
	\frac{i}{6} \epsilon_{\psi \psi}{}^{\ha \hb \hc} \D_s s^i \Sigma_{\ha \hb} \D^j H_{\hc ij} \= \frac{1}{10} s^i \Sigma_{\psi \psi} s^j \left( 2 \partial_\ha H^\ha_{ij} - i \D^2_{\ha ki} H^{\ha k}_j \right) + ~\cdots~ \,,
\end{equation}
and so the full constraint reads
\begin{equation}
\label{eq:higher_cons}
	0 \= C_{ij} \= \D^2_{\ha k (i} H^{\ha k}_{j)} + 6i \partial_\ha H^\ha_{ij} \,.
\end{equation}
Finally, by the argument given at the end of \S\ref{sec:5d_2form_jank}, the top two conditions imply no new constraints on the field-strength inside $H$.

\vspace{2pt}
\thoughtbreak
\vspace{-4pt}

Note that the way this procedure works is not to methodically derive the constraints through careful algebra. The key is to recognize precisely what form the constraint term \textit{must} take and to pull out all the pieces that have that structure. Everything else will either cancel or be absorbed into the next component, as \textit{required} to happen by the logic leading up to \eqref{eq:Bcons_jank}. Throughout the next two subsections we will examine this in a more sophisticated mathematical setting; in \S\ref{sec:triv_coh} we discuss the idea that no new constraints are implied by the top two Bianchi identities in any superspace and in \S\ref{sec:geo_to_alg} we review the way in which we have reduced superform analysis from a super-geometric problem to simple linear algebra.

\subsection{Proof of Trivial Cohomology (\texorpdfstring{$s \leq 1$}{s ≤ 1})}
\label{sec:triv_coh}

At the end of \S\ref{sec:5d_2form_jank} we mentioned that the structure of the five-dimensional $L$-combination proves the fact that the top two Bianchi identities fail to ever impose new constraints on a superform in $\R^{5 \vert 8}$. It should come as no surprise that similar arguments can be made for all superspaces once their $L$-combinations are known. For example, in six dimensions the $L$-combinations (as derived in \S\ref{sec:lorentz_irred}) are
\begin{equation}
	L^1_{s \psi} ~:=~ \gamma_\psi(s,\, \xi) \qquad \text{and} \qquad L^2_{ss \psi \psi \psi} ~:=~ \gamma_{\psi \psi \psi}(s,\, s)
\end{equation}
for arbitrary spinors $s$ and $\xi$. Both require an $s$ and so a Bianchi component $B_{\psi \ldots \psi}$ will never be able to support a constraint term. Interestingly, $B_{s \psi \ldots \psi}$ can only support a term in the case where $B(A)$ is the Bianchi form for a 1-form as $B_{s \psi} \sim s^i \gamma_\psi C_i$. This same kind of story happens in four dimensions, where new constraints will only ever show up at the $s \leq 1$ levels if the superform under consideration is degree-1. In fact, this is true for every \textit{principal} superspace (cf. \S\ref{sec:lorentz_irred}) due to \eqref{eq:principal_srep}.

Principal superspaces turn out to be those with the maximum spacetime dimension for a given number (2, 4, 8, or 16) of supercharges\footnote{This is only true when there are less than 16 supercharges since $\R^{11 \vert 32}$ is not principal.}, thereby giving all other superspaces under dimensional reduction. As can be seen in \eqref{eq:fierz_split}, the only $L$-combination in a principal superspace with a single $s$-variable is $\gamma_\psi(s,\, \xi)$. This means that upon reduction to a lower-dimensional embedded superspace, all $L$-combinations in the lower-dimensional space must have two $s$-variables since \eqref{eq:principal_srep} will split into two pieces (as in \eqref{eq:principal_split} for the reduction from $\R^{6 \vert 8}$ to $\R^{5 \vert 8}$). Thus, we are led to the following conclusions:
\begin{itemize}
	\item[(1)] In a principal superspace, the set of constraints imposed on $p$-forms by the top two Bianchi identities is trivial for $p > 1$. 
	\item[(2)] In a non-principal superspace, the set of constraints imposed on $p$-forms by the top two Bianchi identities is trivial for all $p$.
	\vspace*{0.25cm}
\end{itemize}

By using nothing more than $L$-combinations and the notion of a principal superspace, we have managed to begin the process of proving general statements about the structure of superforms in arbitrary superspaces. These kinds of results are encouraging and suggest that this truly is a natural and clarifying way to look at the geometry of superspace.

\subsection{From Geometry to Algebra}
\label{sec:geo_to_alg}

Solving Bianchi identities is a fundamentally geometric exercise, and practically, these conditions are often very to difficult to solve. However, we have found that we can recast this messy collection of closure conditions into a systematic series of linear algebra problems. Consider a generic Bianchi form $B(\omega)$ and the contraction operator $\iota_{\gamma(s,\, s)}$, defined such that $\iota_{\gamma(s,\, s)} \omega_\janksubp = \omega_\janksubg$. We denote by $\K$ the vector space of possible terms in the Bianchi components. The first and most important thing to note about $\K$ is that in the decomposition
\begin{equation}
	\K \= \ker(\iota) \oplus \T \,,
\end{equation}
the subspace $\T$ consists entirely of terms that are set to zero in the calculations that determine the form components and constraints. Remember why this is true; when the lower Bianchi components vanish---that is, when the lower-dimensional constraints are imposed---the equation \eqref{eq:Bcons_jank} is implied automatically. More explicitly, any and all effects that lower-dimensional constraints can have on a higher Bianchi component are encapsulated in the equation $\T = 0$. In particular, this means that anything sitting inside the kernel of $\iota$ must be untouched by lower constraints.

Using the fact that $\ker(\iota)$ is graded by dimension, we are allowed to split it as $\ker(\iota) = \bigoplus_d \Z^d$ where each $\Z^d$ is given by
\begin{equation}
	\Z^d \= \B^d \oplus \H_1^d \oplus \cdots \oplus \H_k^d \,.
\end{equation}
The subspace $\B^d$ is where the dimension-$d$ component of $\omega$ sits, while each $\H_\ell^d$ is a linearly independent subspace holding a single dimension-$d$ constraint term and $k$ is the number of linearly independent $L$-combinations in a given superspace. Calculationally, isolating any of $\{\B^d,\, \H_\ell^d\}$ simply requires pulling out the corresponding $L$-combination. In particular, it should be noted that no interference happens between subspaces at the same dimension since the $L$-combinations are linearly independent. Furthermore, the $L$-combinations are also graded by dimension and so
\begin{equation}
	\H_i^a \cap \H_j^b ~=\, \{0\} \qquad \text{for}~ i \neq j ~\text{and}~ a \neq b \,.
\end{equation}
This shows why in \S\ref{sec:prior_cons} we did not check whether \eqref{eq:higher_cons} was implied by \eqref{eq:lower_cons}. In the conventional approach such a check is necessary because \eqref{eq:higher_cons} is accompanied by two other ``constraints" that can be shown to follow directly from \eqref{eq:lower_cons}.\footnote{See the discussion of the five-dimensional ``3-cocycle" in \cite{glr}.} But here we already know that \eqref{eq:higher_cons} \textit{must} be independent simply because of the identity \eqref{eq:Bcons_jank} automatically satisfied by the Bianchi components.

By reformulating the superform analysis in this way we have stripped away every possible non-essential calculation and reduced all remaining work to the simple problem of computing $L$-combinations in a given superspace. The precise structure of the components and constraints is laid bare and the remaining necessary calculations, as we have demonstrated in \S\ref{sec:5d_2form_jank} and \S\ref{sec:prior_cons}, become much faster and more illuminating. Another, more mathematical, presentation of this approach is given in \cite{lin_ran} wherein we explicitly demonstrate how much simpler the computations of \cite{glr} can be made.

\section{Dimensional Reduction}
\label{sec:dim_red}

Let us now move away from the study of individual superforms and look at the relationships between superspaces with the same number of supercharges. The motivation for this line of study is the fact that the similarities between the five-dimensional forms of \S\ref{sec:5d_2form_jank}--\S\ref{sec:prior_cons} and the six-dimensional forms of \cite{alr} are striking. In fact, these similarities are indicative of a much more expansive relationship between all superspaces $\R^{m - 1 \vert n} \hookrightarrow \R^{m \vert n}$. As an example, consider the generic Bianchi identity \eqref{eq:jank_bi} for a closed $p$-form $\omega$ in flat six-dimensional, $N = (1,\, 0)$ superspace. Written in $5 + 1$ dimensions, this condition splits into
\begin{align}
\label{eq:red_bi_vec}
	0 & \= s \D_s \omega_\janksubp + (-1)^s (p + 1 - s) \partial_\psi \omega_\janksubp \notag\\
		& \quad + i (-1)^s s(s - 1) \omega_\janksubG - i (-1)^s s (s - 1) c_{ss} \omega_\janksub{6} \,, \\[8pt]
\label{eq:red_bi_6}
	0 & \= s \D_s \omega_\janksub{6} + (-1)^s \partial_6 \omega_\janksubp + (-1)^s (p - s) \partial_\psi \omega_\janksub{6} \notag\\
		& \quad + i (-1)^s s(s - 1) \omega_\janksub{\Gamma(s,\, s) 6} \,,
\end{align}
where $c_{ss} = s^2$ and we have divided a factor of $(p + 1 - s)$ out of the second equation. We now want to look for purely five-dimensional superforms, so let us define $\beta_{p - 1}$ as $\beta_\janksubp := \omega_\janksub{6}$ and $\alpha_p$ as $\omega_p$ with all vector indices restricted to 5D and no $x^6$ dependence. Since the $\partial_6$ term then drops out of \eqref{eq:red_bi_6}, the two equations give
\begin{equation}
\label{eq:dim_red_rel_coh}
	\d \alpha_p \= c_2 \wedge \beta_{p - 1} \,,
\end{equation}
\begin{equation}
	\d \beta_{p - 1} \= 0 \qquad \Rightarrow \qquad \beta_{p - 1} \= \d \theta_{p - 2} \,.
\end{equation}
Notice that only $\beta$ is closed; these are the forms from six dimensions that give rise to the five-dimensional super-de Rham complex. For an explicit illustration of this relationship, refer to the table on the following page where we collect the six-dimensional $p$-forms of \cite{alr} for $p = 2,\, 3,\, 4,\, 5$ and the five-dimensional $p$-forms of \cite{glr} for $p = 1,\, 2,\, 3,\, 4$. We omit real prefactors and use $\star$ to denote the Hodge dual.

\afterpage{
\includegraphics[scale = 1.0, trim = 3.9cm 0 0 3.2cm]{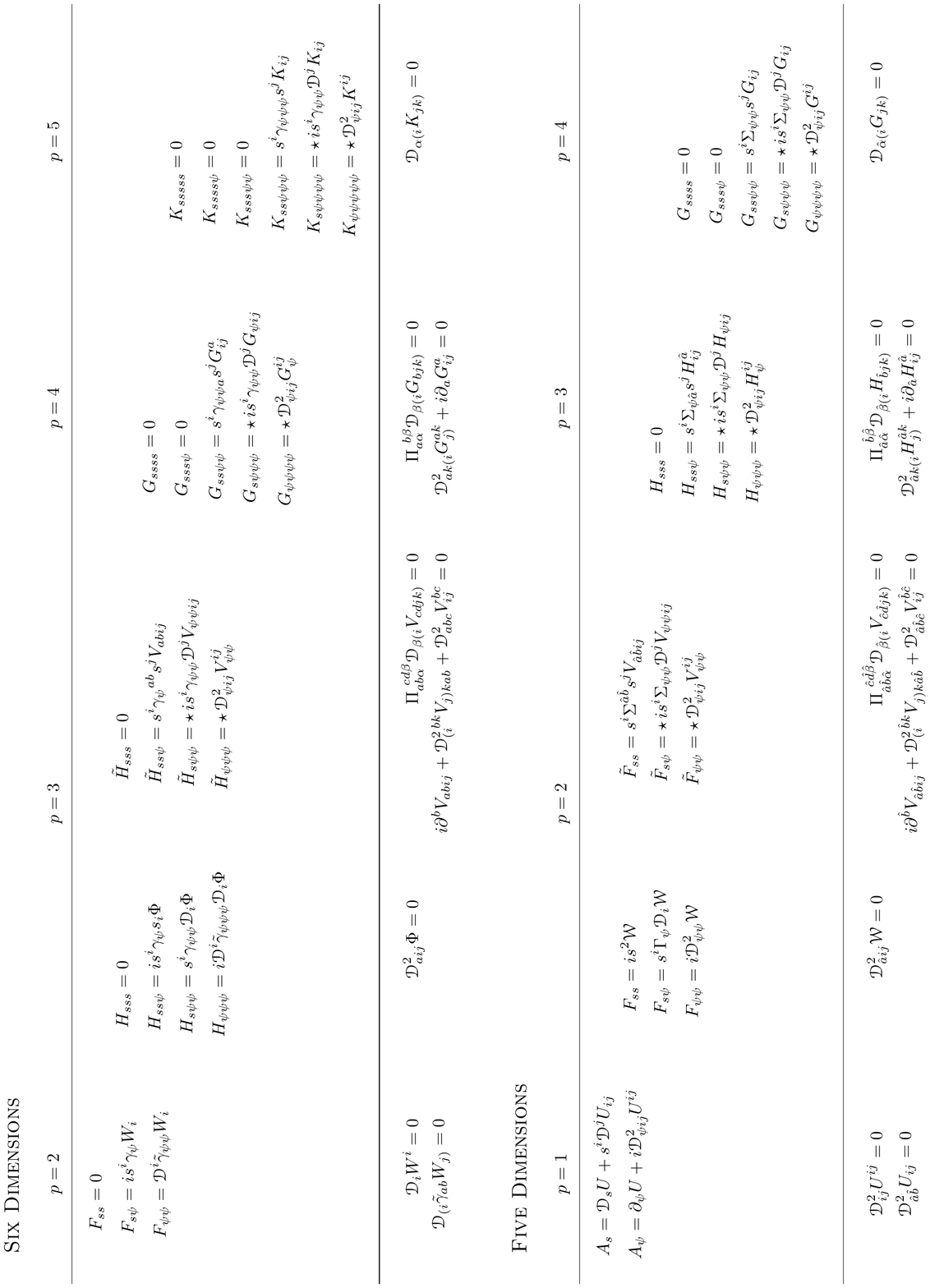}
\thispagestyle{empty}}

Due to the index structure of the fields and the constraints they satisfy, it is very useful to note that the solutions to the constraints reduce appropriately as well. For example, the six-dimensional 4-form constraints are solved by $G_{a ij} = \D^2_{a ij} \varphi$ for some unconstrained superfield $\varphi$. In five dimensions we can do the same with $H_{\ha ij}$ except that the dimension of $\varphi$ gets shifted down by one. The reason for this similarity is simple: in 6D, writing $G_{a ij} = \D^2_{a ij} \varphi$ is equivalent to writing $G = \d H$ and then requiring that $G_{a ij}$ solve the constraints placed upon itself by demanding that $\d G = 0$. However, this is of course already satisfied since $\d^2 = 0$. So because the structure of the complex remains essentially unchanged after dimensional reduction, the identity $\d^2 = 0$ ensures that solutions to the six-dimensional $p$-form constraints, when reduced, are also solutions to the five-dimensional $(p - 1)$-form constraints.

Even more work can be avoided by noting that the reduction of Bianchi identities extends to curved superspaces. This is again because the curved-space torsions involve contractions over (composite) spinor indices and we therefore do not get any new ``broken" pieces like we did with the $c_{ss}$ term. So although we do not explicitly collect the curved superforms in \cite{glr}, it would be a fairly straightfoward task to reduce them from their six-dimensional counterparts in \cite{alr}.

\subsection{Relative Cohomology}
\label{sec:rel_coh}

Returning to the $\alpha_p$ part of the reduction, note that it is possible to construct a closed 5D $p$-form by defining the shifted superform
\begin{equation}
	\alpha'_p ~:=~ \alpha_p - c_2 \wedge \theta_{p - 2} \,,
\end{equation}
which is illustrated here.
\vspace{0.3cm}

\begin{figure}[htb]
\begin{center}
\hspace*{0.75cm}
\begin{tikzpicture}[rotate = 90, scale = 2]
	\draw [color = black!50!white] (-0.65, 2) -- (0.65, 2);
	\draw [color = black!50!white] (-0.65, 1.2) -- (0.65, 1.2);

	\draw (-0.5, 3) -- (-0.5, 1.85);
	\draw (-0.505, 1.785) node[right] {$\cdots$};
	\draw (-0.5, 1.35) -- (-0.5, 0.55);
	\draw (-0.505, 0.485) node[right] {$\cdots$};
	
	\draw (0.5, 3) -- (0.5, 1.85);
	\draw (0.495, 1.785) node[right] {$\cdots$};
	\draw (0.5, 1.35) -- (0.5, 0.55);
	\draw (0.495, 0.485) node[right] {$\cdots$};
	
	\draw [fill = black] (-0.5, 3) circle (0.05);
	\draw [fill = black] (-0.5, 2.5) circle (0.05);
	\draw [fill = white] (-0.5, 0.7) circle (0.05);
	\draw (-0.52, -0.125) node[right] {$\alpha_p$};
	\draw (0.52, -0.125) node[right] {$c_2 \wedge \theta_{p - 2}$};
	
	\draw [fill = white] (0.5, 3) circle (0.05);
	\draw [fill = white] (0.5, 2.5) circle (0.05);
	\draw [fill = black] (0.5, 0.7) circle (0.05);
	
	\draw [fill = black] (-0.5, 2) circle (0.05);
	\draw [fill = black] (-0.5, 1.2) circle (0.05);
	\draw [fill = black] (0.5, 2) circle (0.05);
	\draw [fill = black] (0.5, 1.2) circle (0.05);
\end{tikzpicture}
\label{fig:rel_coh}
\vspace*{0.75cm}
\begin{minipage}{0.75\textwidth}
{\footnotesize Filled nodes are the non-zero components of the indicated forms and the struts denote which components of $\alpha_p$ are ``corrected" by $c_2 \wedge \theta_{p - 2}$ to allow the form $\alpha'_p$ to close without vanishing. The components with higher mass-dimension are further to the left.}
\end{minipage}
\end{center}
\end{figure}
\vspace{-0.3cm}

\noindent Although this form is not what we are usually looking for in terms of forms built from single superfields, $\alpha'_p$ is interesting because it is exactly the form that comes from the relative cohomology construction of a closed 5-form in \cite{hpsc}. The fact that their $L_6 = c_2 \wedge G_4$ exhibits Weil triviality as $L_6 = \d K_5$ and $L_6 = c_2 \wedge \d h_3$ is then a direct consequence of the fact that $G_4$ and $K_5$ come to 5D together as a relative cohomology pair from the dimensional reduction of the 6D 5-form.

The lowest interesting case of relative cohomology comes from the six-dimensional 3-form.\footnote{The relative 2-form turns out to be equivalent (up to zero-mode) to the de Rham 2-form $F(\W)$.} The closed form $H'$ arising from this construction has the components
\vspace*{-0.1cm}
\begin{align}
	H'_{sss} & \= - s^2 A_s \,, \notag\\[2pt]
	H'_{ss \psi} & \= s^i \Gamma_\psi s_i \Phi - s^2 A_\psi \,, \notag\\[2pt]
	H'_{s \psi \psi} & \= \frac{i}{4} s^i \Sigma_{\psi \psi} \D_i \Phi \,, \notag\\[2pt]
	H'_{\psi \psi \psi} & \= \frac{3}{8} \D^2_{\psi \psi \psi} \Phi \,,
\end{align}
where the relationships $\Phi = \tfrac{i}{24} \D^{\ah i} A_{\ah i}$ and $A_\psi = - \tfrac{i}{24} \D^i \Gamma_\psi A_i$ are fixed by the dimension-2 Bianchi identity. Notice how the 1-form $A$ simply helps get the form ``off the ground" by allowing for the non-trivial satisfaction of the lower Bianchi identities before getting out of the way at the higher levels. The constraints imposed by $\d H' = 0$ are
\vspace*{-0.2cm}
\begin{align}
	\label{eq:rc_cons_dim2}
	\D_{(\ah (i} A_{\bh) j)} & \= 0 \,, \\[2pt]
	\label{eq:rc_cons_dim52}
	6 (\Gamma_\ha)_\ah{}^\bh \D_{\bh i} \Phi + 3 (\Sigma_{\ha \hb})_\ah{}^\bh \D_{\bh i} A^\hb - (\Sigma_{\ha \hb})_\ah{}^\bh \partial^\hb A_{\bh i} & \= 0 \,, \\[4pt]
	\label{eq:rc_cons_dim3}
	\D^2_{ij} \Phi & \= 0  \,.
\end{align}
The lower components of this form are not gauge-invariant under the transformation $A_{\ah i} \mapsto A'_{\ah i} + \D_{\ah i} \Lambda$ for some gauge parameter $\Lambda$, although the constraints and top two components are invariant. This is a generic feature of the relative cohomology construction since the lower components of $\alpha'_p$ are being shifted by the $(p - 2)$-form potential $\theta_{p - 2}$ that solves the closure condition $\d \beta_{p - 1} = 0$ as $\beta_{p - 1} = \d \theta_{p - 2}$.

The relative cohomology forms are also more important than they might seem at first glance. As discussed in \cite{glr}, the superform $H$ does not, surprisingly, describe a matter multiplet. Instead, its field content is a collection of superconformal gauge parameters. This is discussed in more depth in \cite{glr} but the real question is then: \textit{where is the tensor multiplet in five dimensions?} The simplest six-dimensional tensor multiplet is outlined in \cite{alr} as the multiplet defined by $\D^2_{a ij} \Phi = 0$. In five dimensions, that constraint would split into a 1-form piece and a scalar piece. As it turns out, the constraints \eqref{eq:rc_cons_dim2} and \eqref{eq:rc_cons_dim52} combine to give $\D^2_{\ha ij} \Phi = 0$. Combined with \eqref{eq:rc_cons_dim3}, this is precisely the on-shell tensor multiplet we expected in 5D. Yet it was the relative cohomology construction $H'$ that produced it, with $H$ playing a very different role.

This is a somewhat surprising conclusion; the conventional wisdom is that supersymmetrizing a bosonic $p$-form is the same as obtaining a $p$-form from the super-de Rham complex. However, even when this was originally laid out in \cite{gates} an extra assumption was made that hid the counterexample to this statement. For the vector multiplet in 4D, $N = 1$ superspace there are two constraints required in order for the field-strength to be a closed superform: the scalar constraint
\begin{equation}
	\D^\alpha W_\alpha - \lbar{\D}_\ad \lbar{W}^\ad \= 0
\end{equation}
and the ``chirality constraint"
\begin{equation}
\label{eq:4d_ch_cons}
	\D_\alpha \lbar{W}_\ad - \lbar{\D}_\ad W_\alpha \= 0 \,.
\end{equation}
The usual choice is to obstruct the first with a scalar superfield $G$ that then describes a tensor multiplet and choose $W_\alpha$ to be chiral to satisfy \eqref{eq:4d_ch_cons}. However, if we had instead chosen to obstruct the second as $\delta H_{\alpha \ad} = \D_\alpha \lbar{L}_\ad - \lbar{\D}_\ad L_\alpha$, we would have found a multiplet of superconformal gauge parameters \cite{1001} exactly as we do in five dimensions. The novelty now is that in five dimensions this multiplet is unavoidable in the standard super-de Rham complex. However, its presence should not be seen as too surprising given that the same kind of multiplet shows up even in four dimensions.

\section{Field Content in 5D}
\label{sec:5d_content}

In \S\ref{sec:bi_usual}, we stated that the utility of closed superforms was in their ability to naturally describe supersymmetric gauge theories. To that end, we now look at three examples of constructing supermultiplets from superform field-strengths: the off-shell vector multiplet, the on-shell tensor multiplet, and the off-shell linear multiplet.

\subsection{The Vector Multiplet (\texorpdfstring{$p = 2$}{p = 2})}
\label{sec:vect_mult}

As derived in \S\ref{sec:5d_2form_vanilla} and again in \S\ref{sec:5d_2form_jank}, the theory of a closed, five-dimensional 2-form has at its core a dimension-1 field-strength $\W$ that satisfies the constraint
\begin{equation}
	\D^2_{\ha ij} \W \= 0 \,,
\end{equation}
which is equivalent to
\begin{equation}
\label{eq:vm_cons}
	\D^{(i}_\ah \D^{j)}_\bh \W \= \frac{1}{4} \epsilon_{\ah \bh} \D^{\gh (i} \D_\gh^{j)} \W \,.
\end{equation}
This constraint, as we will see, defines a five-dimensional vector multiplet \cite{how_lin, kuz_lin}.

Before delving into components and counting degrees of freedom, there are two important things to note. The first is that
\begin{equation}
\label{eq:vm_symm}
	\D^{(i}_\ah \D^{j)}_\bh \W \= \frac{1}{4} \epsilon_{\ah \bh} \D^{\gh (i} \D_\gh^{j)} \W ~~\Rightarrow~~ \D_\ah^{(i} \D_\bh^j \D_\gh^{k)} \W \= 0 \,.
\end{equation}
This will be used later when we look at the degrees of freedom in this multiplet. The second thing to note is that acting on \eqref{eq:vm_cons} with $\D_i^\ah$ yields for the spinor $\lambda$ in $\W$,
\begin{equation}
\label{eq:vec_mult_dirac}
	\spartial_\ah{}^\bh \lambda_{\bh i} \= - \frac{i}{2} \D^2_{ij} \lambda_\ah^j ~\neq~ 0 \,.
\end{equation}
Thus, this multiplet is off-shell. This may seem odd given that the six-dimensional 3-form from which this reduces is on-shell but note that the obstruction to the Dirac equation in \eqref{eq:vec_mult_dirac} is an operator that does not exist in six dimensions.\footnote{Generally, on-shell spinors in $d = 2k$ are off-shell in $d = 2k - 1$ with $\partial_{2k}$ as a central charge.}

Turning now to the field content, we write the $\theta$-expansion of $\W$ as \cite{kuz_lin}
\begin{equation}
	\W \= \phi + i \theta^{\ah i} \lambda_{\ah i} + \frac{i}{2} \theta^{\ah i} \theta_\ah^j X_{i j} + i \theta^{\ah i} \theta^\bh_i F_{\ah \bh} + \mathcal{O}(\theta^3) \,.
\end{equation}
The degrees of freedom in $\W$ are then

\begin{equation}
\label{tab:vm_fields}
\begin{tabular}{c|c|c|c|c}
	fields & $\phi$ & $\lambda^\ah_i$ & $X_{ij}$ & $F^{\ah \bh}$ \\
	\hline \hline
	on-shell & ~1~ & 4 & 0 & 3 \\
	\hline
	off-shell & 1 & 8 & 3 & 4 \\
\end{tabular}
\end{equation}
\vspace{0pt}

\noindent since $F_{\ha \hb} = (\Sigma_{\ha \hb})^{\ah \bh} F_{\ah \bh} = - \tfrac{i}{2} \D^2_{\ha \hb} \W$ and is the field-strength of a dynamical vector\footnote{The counting for such an object is $D - 1$ degrees of freedom off-shell and $D - 2$ on-shell.} due to the dimension-3 Bianchi identity \eqref{eq:2form_dim3_bi}. In order to determine the on-shell degrees of freedom for the iso-triplet $X_{ij}$, we first need to know whether there are any new fields at higher order in $\theta$. To do so, we use the dimension-$\tfrac{5}{2}$ Bianchi identity \eqref{eq:2form_dim52_bi} and consider what components might live in $\D \D \D \W$. To wit, suppose $\D \D \D$ were totally anti-symmetric in spinor indices. If not totally symmetric in isospin, the anti-symmetric spinor + anti-symmetric isospin components would form partial derivatives. However, if it were totally symmetric in isospin, then it would vanish by \eqref{eq:vm_symm}. Therefore the only possible remaining source of new components is $\D \D \D$ with at least one symmetric pair of spinor indices. But these are exactly the terms that \eqref{eq:2form_dim52_bi} rules out. Thus, the fields laid out in \eqref{tab:vm_fields} are the only ones to be found and higher components are simply derivatives of the lower ones.

Coupling this argument with the requirement that supersymmetry hold on-shell, the field $X_{ij}$ is relegated to the role of auxiliary field and cannot carry any on-shell degrees of freedom. Due to these considerations, the Lagrangian for this multiplet is
\begin{equation}
	\L ~\sim~ \partial^\ha \phi\, \partial_\ha \phi + i \lambda^i \spartial \lambda_i + X^{ij} X_{ij} + F^{\ha \hb} F_{\ha \hb} + \lambda^i [\phi,\, \lambda_i] \,.
	\vspace*{0.5cm}
\end{equation}
The correct relative coefficients required for $\delta_\text{SUSY} \int \L = 0$ are given in \cite{kuz_lin}.

\subsection{The Tensor Multiplet (\texorpdfstring{$p = 3$}{p = 3})}
\label{sec:3form_mult}

As discussed in \S\ref{sec:rel_coh}, the simplest tensor multiplet in five dimensions is described by the relative cohomology 3-form $H'$ in terms of a dimension-2 scalar superfield $\Phi$ satisfying the constraints
\begin{equation}
	\D^2_{\ha ij} \Phi \= 0 \qquad \text{and} \qquad \D^2_{ij} \Phi \= 0 \,.
\end{equation}
These can be cast together in the more useful form
\begin{equation}
	\D_\ah^{(i} \D_\bh^{j)} \Phi \= 0 \,.
\end{equation}
A constraint of this form was originally given in \cite{soka} to describe a tensor multiplet in six-dimensional $N = (1,\, 0)$ superspace. As we now know, this also comes directly from constructing the six-dimensional super-de Rham complex as in \cite{alr}.

It is straightforward to check that the $\theta$-expansion of $\Phi$,
\begin{equation}
	\Phi \= \phi + \theta^\ah_i \chi^i_\ah + \theta^{\ah i} \theta_i^\bh T_{\ah \bh} + \mathcal{O}(\theta^3) \,,
\end{equation}
stops giving new fields beyond the $\theta^2$-level. Unfortunately, this means that the multiplet is on-shell with the degrees of freedom

\begin{equation}
\label{tab:tm_fields}
\begin{tabular}{c|c|c|c}
	fields & $\phi$ & $\chi^\ah_i$ & $T^{\ah \bh}$ \\
	\hline \hline
	on-shell & ~1~ & 4 & 3
\end{tabular}
\end{equation}
\vspace{0pt}

\noindent where $T_{\ah \bh} \= (\Sigma^{\ha \hb})_{\ah \bh} T_{\ha \hb}$ is dual to the 3-form field-strength of a 2-form gauge field. Determining how a superform describing an \textit{off-shell} tensor multiplet would sit in $\R^{5 \vert 8}$ is an active topic of research, especially because of its relation to the superspace analogue of the Perry-Schwarz construction \cite{per_sch}.

\subsection{The Linear Multiplet (\texorpdfstring{$p = 4$}{p = 4})}

The supermultiplet content described by a closed, five-dimensional 4-form is contained inside a superfield $G_{ij}$ subject to the analyticity constraint
\begin{equation}
\label{eq:4form_mult_cons}
	\D_{\ah (i} G_{jk)} \= 0 \,.
\end{equation}
This is the five-dimensional $N = 1$ linear multiplet, the four-dimensional $N = 2$ version\footnote{A five-dimensional formulation is given in \cite{bkn} but they do not examine the field content before reducing to a centrally-extended 4D, $N = 2$ superspace.} of which was discovered in \cite{sohnius_lm}. The $\theta$-expansion is
\begin{equation}
	G_{ij} \= \varphi_{ij} + \theta_{(i} \psi_{j)} + \theta_i \Gamma^\ha \theta_j V_\ha + \theta_i \theta_j M + \text{derivatives} \,.
\end{equation}
Additionally, the constraint \eqref{eq:4form_mult_cons} requires that $\partial_\ha V^\ha = 0$. This condition can be solved as
\begin{equation}
	V^\ha \= \epsilon^{\ha \hb \hc \hd \he} \partial_\hb E_{\hc \hd \he}
\end{equation}
for a gauge 3-form $E$. The degrees of freedom carried by these fields are

\begin{equation}
\label{tab:4form_fields}
\begin{tabular}{c|c|c|c|c}
	fields & $\varphi_{ij}$ & $\psi_\ah^i$ & $E^{\ha \hb \hc}$ & $M$ \\
	\hline \hline
	on-shell & 3 & 4 & 1 & 0 \\
	\hline
	off-shell & 3 & 8 & 4 & 1 \\
\end{tabular}
\end{equation}
\vspace{0pt}

\noindent and so the supermultiplet is off-shell. Given this, the action for the linear multiplet must have the form
\begin{equation}
	\L ~\sim~ \partial_\ha \varphi^{ij} \partial^\ha \varphi_{ij} + V^\ha V_\ha + i \psi^i \spartial \psi_i + M^2 \,.
\end{equation}
This is consistent with the four-dimensional textbook treatment given in \cite{gios}, which also holds the correct relative coefficients.

\section{Conclusions}

In this thesis we have introduced a new approach to the study of closed superforms. We showed how this approach isolates constraints and easily allows for a proof of the statement that the top two closure conditions almost never impose any new constraints on the field-strength. We also discussed the general process of dimensional reduction, giving $\R^{6 \vert 8} \rightarrow \R^{5 \vert 8}$ as an example, and noted, through our introduction to the relative cohomology construction, that not all gauge supermultiplets can be found in the usual de Rham complex. This is an important point; it is generally not true that the supersymmetrization of the bosonic de Rham complex is equivalent to the de Rham complex of superforms. In four dimensions this inequivalence is avoidable but in higher dimensions it inevitably shows up at certain degrees. Finally, we specialized to five dimensions and wrote down in components the gauge multiplets defined by $p$-form field-strengths for $p = 2,\, 3,\, 4$.

It is our hope that through this thesis we have made the geometry of superspace less mysterious and precisely defined a number of interesting characteristics that may lead to further avenues of fruitful study. In particular, we wish to note that in addition to facilitating the study of gauge theories, this procedure for examining superforms may also give new insights into the structure of supergravity. In the same way that the field-strength constraints were originally derived through arduous tensor calculus computations, the defining supergravity constraints come from difficult calculations involving the higher-dimensional closure conditions. It therefore seems possible to find analogous ``$L$-combinations" for the supergravity torsion constraints that would allow for their simple extraction from the Bianchi identities as well.

However, there are still questions that may require even more tools to answer. For example, one particular open problem is the fact that we still do not know what an off-shell tensor multiplet looks like in $\R^{5 \vert 8}$ or $\R^{6 \vert 8}$ with a finite number of auxiliary fields. For the tensor multiplet in five dimensions, one might think to start with the vector-tensor multiplet of 4D, $N = 2$ centrally-extended superspace \cite{bho, novak}; unfortunately, lifting the vector-tensor multiplet leads to the on-shell tensor $H'$.

This issue is related to the Perry-Schwarz construction \cite{per_sch} mentioned in the introduction that served as the impetus for our systematic study of superforms. Although we are still unable to find the multiplet we want, we now know much more about why such a model is difficult to embed into superspace. The formulation of the 3-form $H'$ \textit{via} dimensional reduction makes it clear that the Perry-Schwarz approach, which only involves the abelian 2-form of the $(2,\, 0)$ theory, has no chance of working in superspace without also breaking apart the representations of the spinors that they neglected but supersymmetry required. Interestingly, this reasoning appears to lead to a description in $\R^{4 + 2 \vert 8}$ where the supersymmetry can be realized off-shell by embedding the fields into a tensor multiplet and two vector multiplets in $\R^{4 + 2 \vert 4}$ that are in turn collected in a pair of $N = 2$ superfields.

As we improve our understanding of dimensional reduction in superspace, the picture for formulating higher-dimensional gauge theories continues to clear. Additionally, it is easy to see why this problem has been so complicated historically; as we noted in five dimensions, the superform construction of a tensor multiplet is not a part of the super-de Rham complex and cannot even be built without explicit reference to the potential! Embedding these gauge theories in superspace may then require knowledge of relative cohomology or even more exotic constructions not explored in this thesis. Regardless of what the correct ideas turn out to be, we are confident that through the development of our new tools and perspectives these obstacles can be overcome. We are better-equipped now than ever before to understand the features of supergravity and supersymmetric gauge theories in full generality.

\end{spacing}

\vspace{12pt}
\noindent \begin{minipage}{0.4685\textwidth}
	\hrulefill
\end{minipage} \hspace{0.01\textwidth}
\begin{minipage}{0.04\textwidth}
	\vspace{-1pt}
	\decosix
\end{minipage} \hspace{-0.012\textwidth}
\begin{minipage}{0.4685\textwidth}
	\hrulefill
\end{minipage}

\clearpage

\appendix

\section{Lorentz-Irreducibles in Principal Superspaces}
\label{sec:lorentz_irred}
\begin{spacing}{1.3}

Let us say that a superspace is \textit{principal} if there exists a spinor representation for that space such that the $\gamma$-matrices satisfy
\begin{equation}
\label{eq:principal_srep}
	\gamma^a(s,\, s) \gamma_a(s,\, \xi) \= 0 \,,
\end{equation}
where $a \in \{0,\, \ldots,\, D - 1\}$ and $s$ and $\xi$ are arbitrary spinors. Of all the superspaces with bosonic dimension $m \leq 11$, the only principal ones are $\R^{3 \vert 2}$, $\R^{4 \vert 4}$, $\R^{6 \vert 8}$, and $\R^{10 \vert 16}$; that is, those with maximal bosonic dimension for 2, 4, 8, or 16 supercharges. Since \eqref{eq:principal_srep} is the sole constraint on the spinor structure of these superspaces, the $L$-combinations \eqref{eq:Lc_jank} must follow from this identity. Furthermore, through dimensional reduction the $L$-combinations in principal superspaces define those in non-principal spaces. Given the role $L$-combinations play in determining the structure of super-de Rham complexes, the condition \eqref{eq:principal_srep} therefore has very deep consequences. In the following subsection, we give an example of how to determine the $L$-combinations in $\R^{5 \vert 8}$ using \eqref{eq:principal_srep} for $\R^{6 \vert 8}$ and dimensional reduction.

\subsection{Five-dimensional \texorpdfstring{$L$}{L}-combinations}

For the purpose of understanding the examples given throughout this thesis we must understand the $L$-combinations in $\R^{5 \vert 8}$ (cf. \S\ref{sec:susy_math} for conventions and notation). The utility of defining principal superspaces comes in giving us a starting point for finding these combinations. Since $\R^{6 \vert 8}$ has the same number of supercharges, let us begin by studying the $L$-combinations there. Aside from \eqref{eq:principal_srep}, we can find any additional combinations by setting $\xi = \tilde{\gamma}_{a_1 \ldots a_p} s$. Then \eqref{eq:principal_srep} becomes
\begin{equation}
\label{eq:fierz_split}
	0 \= s \gamma^{a_0} s s \gamma_{a_0} \tilde{\gamma}_{a_1 \ldots a_p} s \= s \gamma_{[a_1} s s \gamma_{a_2 \ldots a_p]} s + b s \gamma^{a_0} s s \gamma_{a_0 a_1 \ldots a_p} s \,,
\end{equation}
for some irrelevant relative coefficient $b \neq 0$. Thus, we have a new $L$-combination whenever we can get the first term here to vanish. Noting that $\gamma(s,\, s)$ is bosonic, the $p = 2$ case gives
\begin{equation}
\label{eq:principal_gabc}
	\gamma^a(s,\, s) \gamma_{abc}(s,\, s) \= 0 \,.
\end{equation}
Because the $p = 3$ case will clearly not work, \eqref{eq:principal_srep} and \eqref{eq:principal_gabc} define the only $L$-combinations in six dimensions.

Reducing to $\R^{5 \vert 8}$ is then very simple. The defining relation \eqref{eq:principal_srep} becomes
\begin{equation}
\label{eq:principal_split}
	\Gamma^\ha(s,\, s) \Gamma_\ha(s,\, \xi) + s^2 \xi_s \= 0 \,,
\end{equation}
and so $\Gamma(s,\, s)$ now fails to be an $L$-combination. The other condition \eqref{eq:principal_gabc} gives
\begin{equation}
\label{eq:Lc_5}
	\Gamma^\ha(s,\, s) \Sigma_{\ha \hb}(s^i,\, s^j) \= 0
\end{equation}
and
\begin{equation}
	\Gamma^\ha(s,\, s) \Sigma_{\ha \hb \hc}(s^i,\, s^j) + s^2 \Sigma_{\hb \hc}(s^i,\, s^j) \= 0 \,.
\end{equation}
Thus, the sole $L$-combination in $\R^{5 \vert 8}$ is defined by \eqref{eq:Lc_5}.

\section{Five-Dimensional Superspace Mathematics}
\label{sec:susy_math}
\setcounter{equation}{0}

Our five-dimensional notation and conventions were first given in \cite{kuz_lin}. Using the ``mostly-plus" flat metric $\eta_{\ha \hb}$, for $\ha,\, \hb \in \{ 0,\, 1,\, 2,\, 3;\, 5\}$, our $\Gamma$-matrices $\Gamma_\ha = (\Gamma_a,\, \Gamma_5)$, with $a \in \{0,\, 1,\, 2,\, 3\}$, are chosen to satisfy the algebra
\begin{equation}
\label{eq:5d_clifford}
	\{ \Gamma_\ha \,, \Gamma_\hb \} \= - 2 \eta_{\ha \hb} \mathbf{1} \,.
\end{equation}
In order to completely span the space of $4 \times 4$ matrices we introduce the symmetric matrices $\Sigma_{\ha \hb} := - \frac{1}{4} [\Gamma_\ha,\, \Gamma_\hb]$ to complement the anti-symmetric spinor metric $\epsilon_{\ah \bh}$ and anti-symmetric, traceless $\Gamma$-matrices. We also make frequent use of the following identities for $A_{ij} = A_{[ij]}$:
\begin{equation}
	A_{ij} \= \tfrac{1}{2} \epsilon_{ij} A^k{}_k \qquad \text{and} \qquad A^{ij} \= - \tfrac{1}{2} \epsilon^{ij} A^k{}_k \,,
\end{equation}
where $\epsilon_{ij}$ is the isospinor metric. The algebra of 5D, $N = 1$ superspace is then
\begin{equation}
\label{eq:5dN1_alg}
	\{ \D_\ah^i,\, \D_\bh^j \} \= - 2 i \epsilon^{ij} \spartial_{\ah \bh} \,.
\end{equation}
It will be useful to define the irreducible $\D^2$ operators in five dimensions, which are normalized as follows:
\begin{equation}
	\D^2_{ij} ~:=~ \tfrac{1}{2} \D^\ah_{(i} \D_{\ah j)} \,, \qquad \D^2_{\ha ij} ~:=~ \tfrac{1}{2} \D_{(i} \Gamma_\ha \D_{j)} \,, \qquad \D^2_{\ha \hb} ~:=~ \tfrac{1}{2} \D^i \Sigma_{\ha \hb} \D_i \,.
\end{equation}
With these, we can expand a generic $\D \D$ object as
\begin{align}
\label{eq:dd_exp}
	\D_{\ah i} \D_{\bh j} & \= i \epsilon_{ij} \spartial_{\ah \bh} - \tfrac{1}{2} \epsilon_{ij} (\Sigma^{\ha \hb})_{\ah \bh} \D^2_{\ha \hb} + \tfrac{1}{2} \epsilon_{\ah \bh} \D^2_{ij} + \tfrac{1}{2} (\Gamma^\ha)_{\ah \bh} \D^2_{\ha ij} \,.
\end{align}
We also define the shorthand
\begin{equation}
	\D^2_{\ha \hb \hc} ~:=~ - \tfrac{1}{12} \epsilon_{\ha \hb \hc}{}^{\hd \he} \D^2_{\hd \he}
\end{equation}
so that $\epsilon_{\ha \hb}{}^{\hc \hd \he} \D^2_{\hc \hd \he} = \D^2_{\ha \hb}$. Finally, we note the following $\Gamma$-matrix identities that follow directly from \eqref{eq:5d_clifford} as worked out in \cite{kuz_tm}: the completeness relation
\begin{equation}
\label{eq:5d_completeness}
	\epsilon_{\ah \bh \gh \dh} \= \frac{1}{2} (\Gamma^\ha)_{\ah \bh} (\Gamma_\ha)_{\gh \dh} + \tfrac{1}{2} \epsilon_{\ah \bh} \epsilon_{\gh \dh} \,,
\end{equation}
the trace identities
\begin{equation}
\label{eq:gab_trace}
	\operatorname{tr} \Gamma^\ha \Gamma^\hb \= - 4 \eta^{\ha \hb} \qquad \text{and} \qquad \operatorname{tr} \Sigma^{\ha \hb} \Sigma_{\hc \hd} \= - 2\delta^{[\ha}_{[\hc} \delta^{\hb]}_{\hd]} \,,
\end{equation}
and the expansions
\begin{align}
\label{eq:gab_gamma_gamma}
	(\Gamma^\ha)_\ah{}^\gh (\Gamma^\hb)_\gh{}^\bh & \= - \eta^{\ha \hb} \delta_\ah^\bh - 2 (\Sigma^{\ha \hb})_\ah{}^\bh \,, \\[4pt]
\label{eq:gab_gamma_sigma}
	(\Gamma^\ha)_\ah{}^\gh (\Sigma^{\hb \hc})_\gh{}^\bh & \= - \frac{1}{2} \epsilon^{\ha \hb \hc \hd \he} (\Sigma_{\hd \he})_\ah{}^\bh + \eta^{\ha [\hb} (\Gamma^{\hc]})_\ah{}^\bh \,.
\end{align}
\end{spacing}

\clearpage

\renewcommand*{\refname}{\vspace*{-1em}}

\end{document}